\documentclass[prl,twocolumn,showpacs,superscriptaddress]{revtex4-1}

\usepackage{graphicx}
\usepackage{dcolumn}
\usepackage{amsmath}
\usepackage{epsfig}
\usepackage{bm}
\usepackage{amssymb}
\usepackage{float}
\usepackage{natbib}
\usepackage{color}
\usepackage{hyperref}
\hypersetup{colorlinks=true,linkcolor=blue,citecolor=blue,urlcolor=black}

\begin{document}

\title{Quantum-enabled electrometer measures field transients and correlation function}

\author{E.K. Dietsche}
\author{A. Larrouy}
\author{J.M. Raimond}
\author{M. Brune}
\author{S. Gleyzes}
\affiliation{Laboratoire Kastler Brossel, Coll\`ege de France,
  CNRS, ENS-Universit\'e PSL,
  Sorbonne Universit\'e, \\11, place Marcelin Berthelot, 75005 Paris, France}
\email{gleyzes@lkb.ens.fr}

\date{\today}

\begin{abstract}
We realize a non-invasive electrometer based on state engineering in a Rydberg hydrogenic manifold. A quantum interference process involving states with very different dipoles measures directly the time correlation of a stochastic electric field, with a 200~ns time resolution and with a $83.2\pm1.4$~mV/m single-atom sensitivity, beyond the Standard Quantum Limit in this context. This quantum-enabled correlation electrometer opens the way to applications in mesoscopic physics, e.g. for the study of individual charge transits in nano-structures.
\end{abstract}

\maketitle

Recent progresses in mesoscopic physics make it possible 
to directly observe the transit of individual charges through nano-structures and record the statistics of the current fluctuations~\cite{Lu2003,Bylander2005,Fujisawa2006,Gustavsson2006}, which provides much more information than the average current~\cite{Blanter2000}. This requires utterly sensitive electrometers~\cite{Jalil2017}. A wide variety of solid-state devices have been proposed and realized, ranging from single-electron transistors~\cite{Yoo1997,Schoelkopf1998} or electromechanical resonators~\cite{Cleland1998,Bunch2007} to sensors based on nitrogen-vacancy centers~\cite{Dolde2014} or semiconductor quantum dots~\cite{Vamivakas2011,Houel2012,Arnold2014}. 

An alternative route for high-sensitivity single-charge detection relies on Rydberg atoms (excited states with a high principal quantum number $n$~\cite{Gallagher1994}), strongly coupled to the electric field due to their huge dipoles. We demonstrated recently a non-invasive ultra-sensitive Rydberg-atom electrometer~\cite{Facon2016}. The field is measured by a quantum interference process involving two states with very different dipoles. The sensitivity is beyond the Standard Quantum Limit (SQL) in this context. The field could be sampled on a small time interval, but the limited repetition rate of the sequence prevented us  from characterizing high-frequency stochastic fields.

In this Letter, we overcome this limitation and demonstrate the operation of a Rydberg-based high-sensitivity electrometer measuring the electric field correlation function with a 5 MHz bandwidth. It measures the spectral density of a stochastic electric field, in the spirit of noise spectroscopy~\cite{Bylander2011,Alvarez2011,Dial2013,Romach2015} or quantum lock-in experiments~\cite{Kotler2011}. This determination is essential to tailor dynamical decoupling protocols improving the electrometer sensitivity in a selected frequency range~\cite{BarGill2012}. We also envision interesting applications in mesoscopic physics, to measure single charges time correlation functions.

The Stark levels of Rubidium  (principal quantum number $n=51$, in an electric field $\mathbf{F}$ aligned along the quantization axis $Oz$) are represented schematically on Fig.~1(a). They are sorted according to their magnetic quantum number $m$ (we only represent  those with $m\ge0$). They arrange in a triangular array, with the circular state $nC$ ($m=n-1$)~\cite{Hulet1983} at its tip [green level on Fig.~1(a)]. For $m\ge 3$, the quantum defects due to the finite size of the Rubidium ion core are negligible and the levels are hydrogenic. For $m\le 2$ the quantum defects appreciably shift the Stark levels from their hydrogenic positions. Neglecting the second-order Stark effect, all dipole-allowed transitions between adjacent levels with $m>2$ in the hydrogenic manifold occur at the Stark frequency $\omega_n=(3/2) n ea_0 F$, where $a_0$ is the Bohr radius ($\omega_{51}/2\pi=98$~MHz for  $F=1$~V/cm).  

The circular state $|51C\rangle$ can be prepared by laser and radio-frequency (rf) excitation in an efficient, selective and fast process~\cite{Signoles2017}. From this initial state, we explore selectively the lower edge of the Stark structure [blue levels in Fig. 1(a)] by shining a  rf field, resonant at $\omega_{51}$, with a pure $\sigma^+$ polarization. This ladder of equidistant states is equivalent to a large angular momentum  (spin) $J_1=(n-1)/2$. A resonant rf pulse performs a rotation by an angle $\theta_1$  of this spin on its Bloch sphere (the circular state is mapped on the North pole) and prepares a  Spin Coherent State (SCS), as long as we remain far enough from the non-hydrogenic region with $m\le 2$.  Close to the South pole, the SCS is a superposition of states with large positive dipoles, proportional to $n^2$ i.e. to the size of the Rydberg orbit. Similarly, a  $\sigma^-$-polarized resonant rf pulse addresses selectively the higher edge of the manifold [orange levels in Fig. 1(a)], equivalent to a spin $J_2=(n-1)/2$, evolving on a second Bloch sphere.  A rf-induced rotation  by an angle $\theta_2$ from $|51C\rangle$ prepare a SCS of spin $J_2$, superposition of states with large negative dipoles. 

\begin{figure}
\includegraphics[width=.9\linewidth]{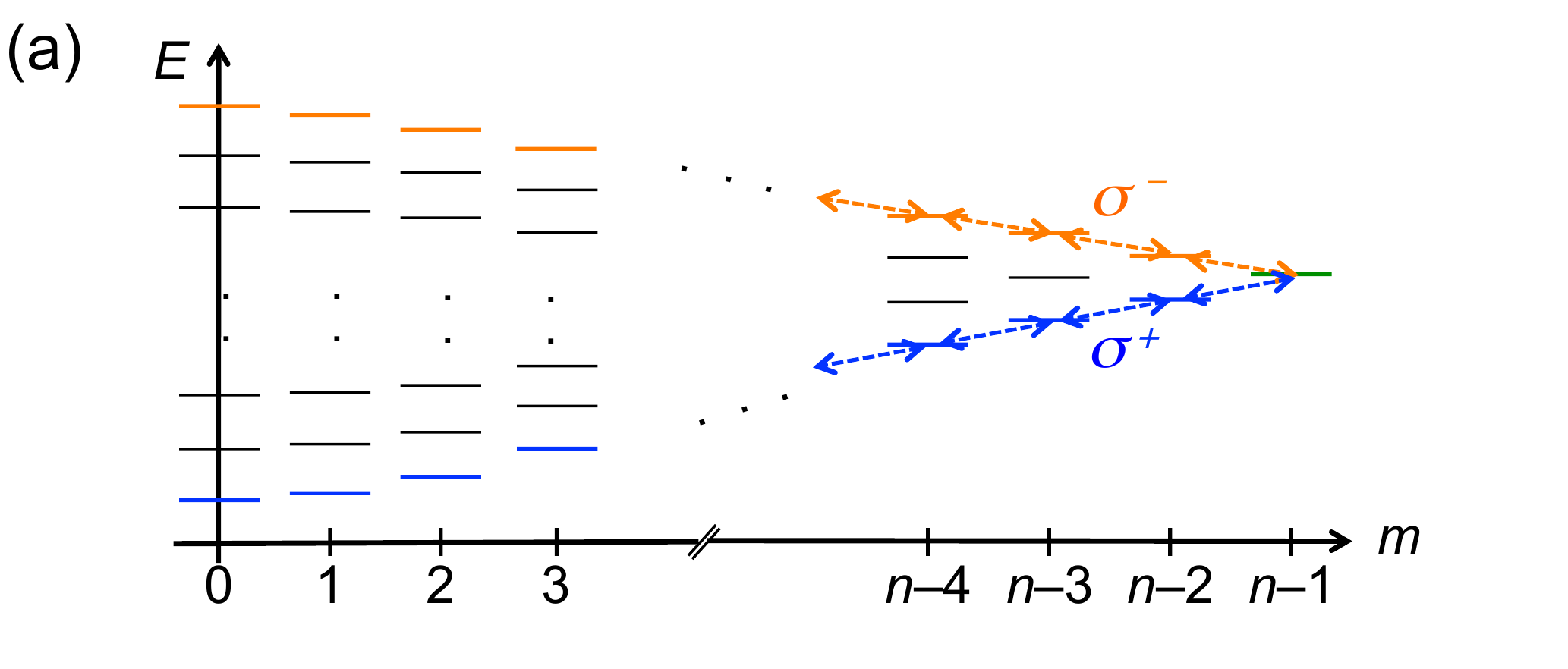}
\includegraphics[width=.9\linewidth]{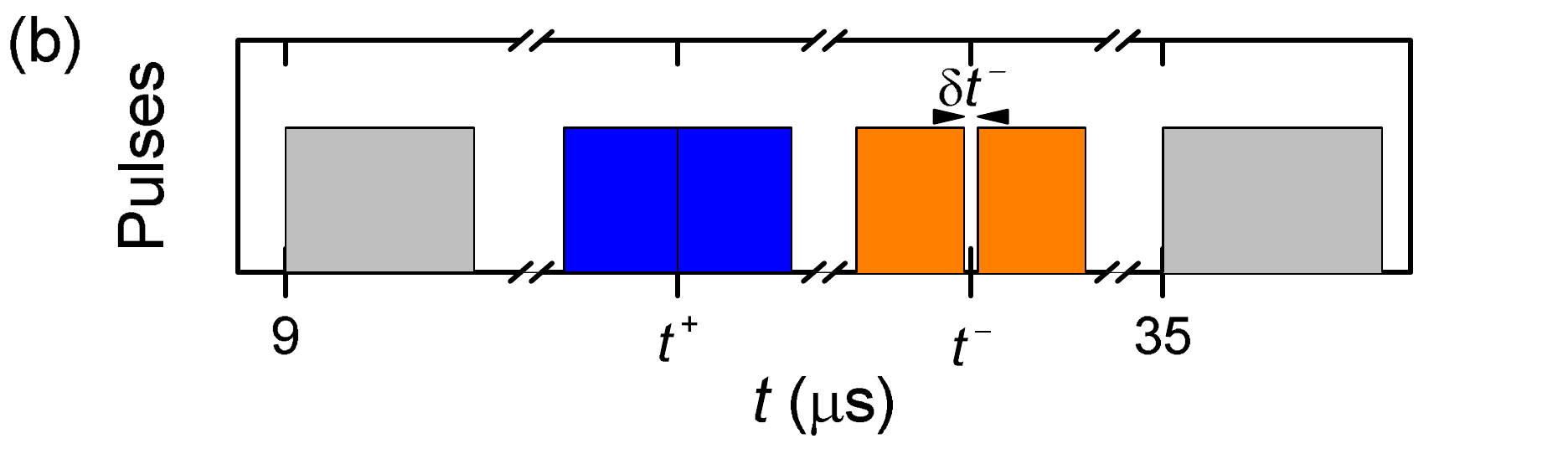}
\caption{ (a) Energy diagram of the $n=51$ manifold of the Rubidium atom in an electric field (not to scale). The levels with $m\ge0$, sorted by their magnetic quantum number, form a triangular structure. The  levels with $m\ge3$ are hydrogenic and equally spaced. For $m\le2$, the level are appreciably shifted from the hydrogen-like positions. A $\sigma^+$( $\sigma^-$)-polarized  rf field couples the circular state (in green) to the lowest (highest) levels for each value of $m$, represented in blue (orange). (b) Timing of the experiment. Gray blocks:  microwave pulses creating and probing a superposition of $|51C\rangle$ and $|49C\rangle$. Blue (orange) blocks: $\sigma^+$ ($\sigma^-$) rf pulses creating a SCS with a large positive (negative) dipole around  $t=t^+$ ($t=t^-$). }
\label{fig-1}
\end{figure}

Our correlation electrometer relies on a sequential exploration of two SCSs with opposite dipoles. We engineer a state that has, at one time, a high positive permanent dipole and, at another time, a high negative one. A quantum interference process between this engineered state and a reference state, with a small polarizability, provides information on the difference between electric field values at different times. 

The experimental set-up is described in details in~\cite{Facon2016,Signoles2017}.
 A thermal Rubidium atomic beam is sent across a plane-parallel capacitor, generating the electric field $\mathbf{F}$. It is the sum of a large time-independent field, $F_0=2.345$~V/cm, defining the quantization axis $Oz$, and of the small field, $f(t)$, to be measured: $F(t)=F_0+f(t)$ ($|f|\ll F_0$). Radio-frequency potentials, carefully tuned in amplitude and phase, applied on four electrodes located on the side of this condenser, generate, close to the center of the structure, a rf field with a nearly pure $\sigma^+$  or $\sigma^-$ polarization. At time $t=0$, the 
Rydberg states are prepared by pulsed  (1~$\mu$s) laser excitation of the ground-state beam, followed by a transfer into $|51C\rangle$ ending at $t=8.1$~$\mu$s. We prepare, on the average, about one atom in $|51C\rangle$ inside the $\approx 1~$mm$^3$ excitation region, with a  $252\pm 7$~m/s velocity, selected in the Doppler profile by the exciting lasers. This atom finally reaches, outside the capacitor, after at $\approx$ 300~$\mu$s flight, a field-ionization detector $D$ resolving adjacent Rydberg manifolds. 

Fig. 1(b) (upper frame) presents the experimental sequence timing. At time $t=9\ \mu$s, a 0.5~$\mu$s $\pi/2$ microwave (mw) pulse at 52.678~GHz, resonant with the two-photon $51C\rightarrow49C$ transition, prepares the state $|\Psi(0)\rangle=(|49C\rangle+|51C\rangle)/\sqrt 2$.  The rf field at frequency $\omega_{rf} = 2\pi \cdot 230 $ MHz $\approx \omega_{51}$ is resonant with the Stark frequency in the $n=51$ manifold (the small signal field $f(t)$ does not appreciably change the rf resonance frequency).  With $F_0=2.345$~V/cm, $(\omega_{51}-\omega_{49})/2\pi=9$~MHz. Thus, the $n=49$ manifold is not resonant with the rf and, in first approximation, is impervious to it \cite{supplementary}. The $|49C\rangle$ state is a reference used to measure, by Ramsey interferometry, the quantum phase accumulated by the part of the wavefunction in the $51$ manifold. 

After the first mw pulse, we apply a set of two consecutive $t_{rf}=102$~ns-long $\sigma^+$-polarized  rf pulses centered on time $t^+$. The first performs a rotation of spin $J_1$ by an angle $\theta_1\simeq144^\circ$. The phase of the second pulse is adjusted to bring this SCS back into  $|51C\rangle$. Transiently, the part of the wave-function originally in $|51C\rangle$ acquires a large positive dipole, $D(t)$, which results in the accumulation of a quantum phase $\Phi_+=\int D(t)f(t)\,dt/\hbar$. The atomic state after the two rf pulses is $|\Psi(0)\rangle=(|49C\rangle+\exp(i\Phi_+)|51C\rangle)/\sqrt 2$. The time interval during which $D(t)$ is large, of the order  $t_{rf}$, sets the time resolution of the electrometer. When $f(t)$ is slowly varying at this time scale, we get $\Phi_+\simeq\alpha_+f(t^+)$. We measure $\alpha_+=0.0193\pm0.0001$~rad/(mV/m) \cite{supplementary}. 

After an adjustable delay, we apply a set of two $\sigma^-$ polarized rf pulses ($97$~ns duration), separated by a small time interval $\delta t^-$, centered at time  $t^-$. They bring $|51C\rangle$ into a SCS of spin $J_2$ with a large negative dipole (rotation by an angle $\theta_2\simeq 137^\circ$) and back. The relevant part of the wave-function accumulates a phase $\Phi_-\simeq\alpha_-f(t^-)$ ($\alpha_- <0$). We precisely adjust $\delta t^- = 12$~ns so that $\alpha_-=-\alpha_+=-\alpha$. The atomic state becomes then $|\Psi(0)\rangle=(|49C\rangle+\exp(i\Phi_{tot})|51C\rangle)/\sqrt 2$ with $\Phi_{tot}\simeq\alpha(f(t^+)-f(t^-))$.

Finally, we close the Ramsey interferometer at $t=35\ \mu$s with a second $\pi/2$ mw pulse with an adjustable phase $\varphi_{mw}$ with respect to the first. The final probability, $P_{49}$, to detect the atom in $|49C\rangle$ reads
\begin{equation}
	P_{49}=P_0+\frac{C_0}{2}\cos(\Phi_{tot}-\varphi_{mw})\ ,
	\label{eqn-cos}
\end{equation}
where  the contrast $C_0 = 0.63\pm0.01$ and the offset $P_0=0.43$ slightly depart from their ideal values (1 and $1/2$) due to experimental imperfections.

In order to assess the differential electrometer sensitivity, we analyze its response to a deterministic signal.  We set $\varphi_{mw}=\pi/2$ so that $P_{49}\approx P_0$ when $f(t)=0$. For a small $f(t)$, $P_{49}\approx P_0 + G_0\left(f(t^+)-f(t^-)\right),$ where $G_0=\alpha C_0/2$. We arbitrarily choose  $t^--t^+ = 9\ \mu$s and sample a time-varying signal $f(t)$ made up of two successive $1\ \mu$s pulses, with 100 ns edge times, starting at $t_s$ and $t_s+8.5\ \mu$s, each creating a field $f_0 = 37.6$~mV/m. We scan $t_s$ around $t^+$. The timing corresponding to three values of $t_s-t^+$  are shown in Fig. 2(a).

Fig. 2(b) presents the measured $P_{49}$ values (dots) as a function of $t_s-t^+$, together  with the ideal signal $P_0+(C_0/2)\sin\{\alpha[f(t^+)-f(t^-)]\}$ (red line). For $ t_s-t^+<-1.1\ \mu$s and for $ t_s-t^+>0.6\ \mu$s, $f=0$ during the rf pulses and we accordingly measure $P_{49}=P_0$. In between, we observe three plateaus. Around $t_s-t^+=-0.75\ \mu$s [dotted red line on Fig. 2(b)] the field $f_0$ is applied only during the first pair of rf pulses and $P_{49}>P_0$.  Around $t_s-t^+=0.25\ \mu$s (dotted blue line), the field is applied during the second pair of rf pulses and $P_{49}<P_0$. The sensitivity to the field around $t^-$ is opposite to that around $t^+$ as expected. Finally, around $t_s-t^+=-0.25\ \mu$s (dotted green line), the field $f_0$ is applied during the two rf pulse pairs. We get  $P_{49}=P_0$ even though $f(t^+)=f(t^-)\ne0$, showing that the phases accumulated at $t^+$ and $t^-$ cancel each other. We infer from these data~\cite{supplementary} a single-atom sensitivity $\sigma^1_{\delta \! f} = 82.3 \pm 1.4$~mV/m for the measurement of $\delta \! f =f(t^+)- f(t^-)$. This sensitivity is 2.64(15) dB below the SQL~\cite{Facon2016,supplementary}, demonstrating quantum advantage in this measurement.

The rise and fall times of the experimental signal $P_{49}$ are slightly longer than those (100~ns) of $f(t^+)-f(t^-)$ [inset in Fig. 2(b)], giving a first insight into the frequency response of the electrometer. In order to characterize more precisely this response, we apply around $t^-$ a sinusoidal signal at frequency $\nu$ with an amplitude  $\tilde f(\nu)$ and an adjustable phase $\varphi$ [while $f(t^+)=0$]~\cite{supplementary}. 

\begin{figure}

\includegraphics[width=.9\linewidth]{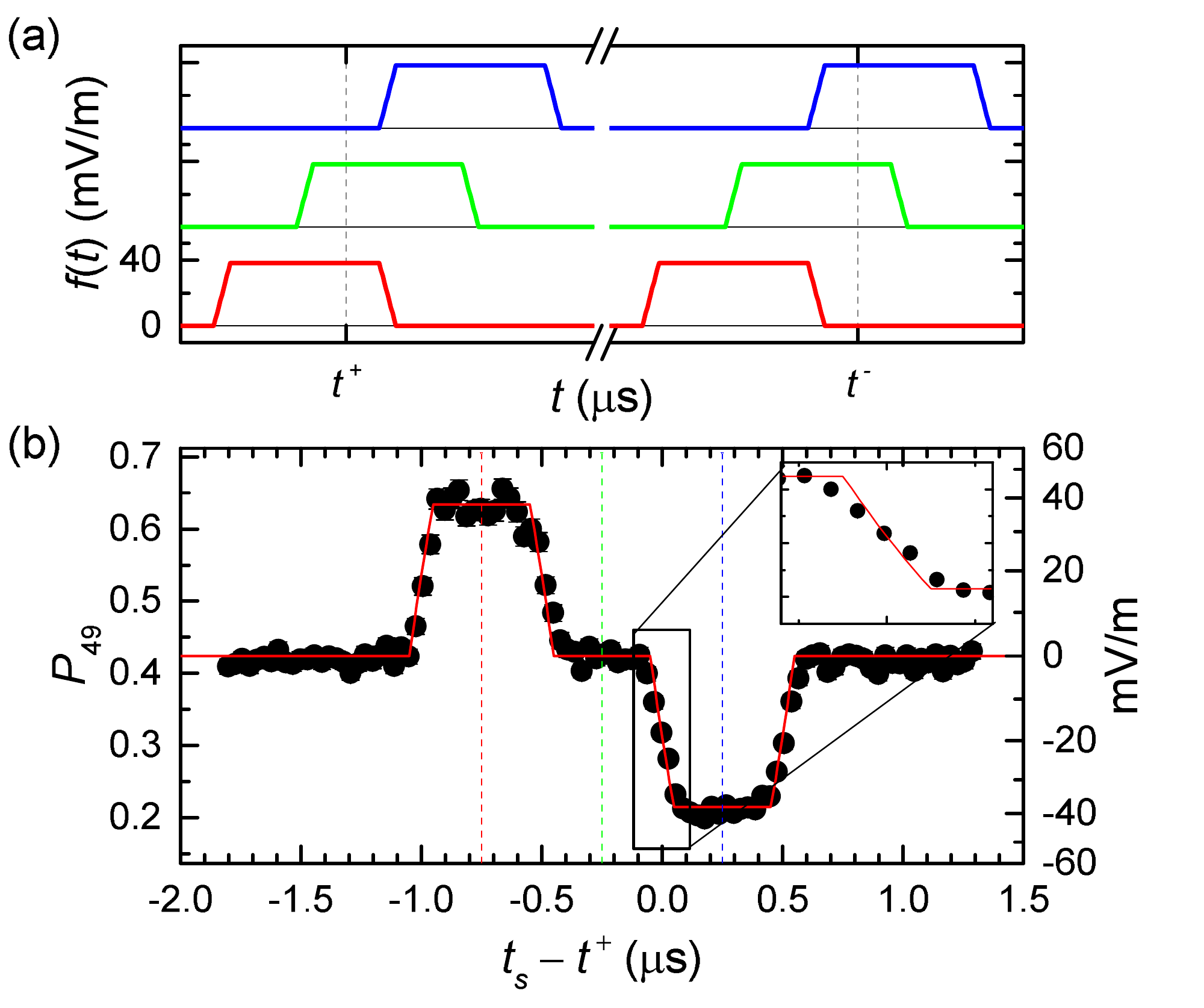}
\vspace{-2mm}

\includegraphics[width=.9\linewidth]{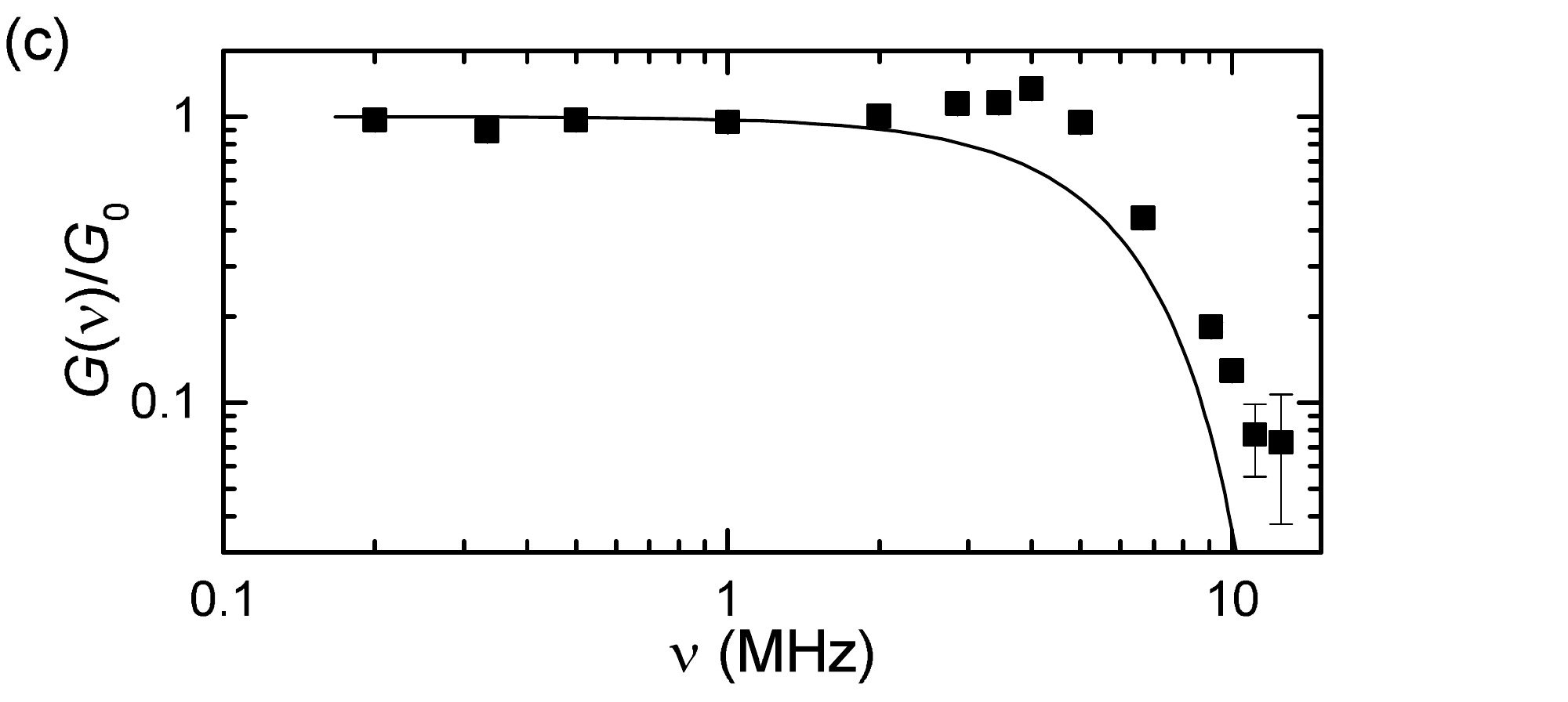}
\caption{ (a) Timing of the electric field pulse for $t_s-t^+=-0.75\ \mu$s (red), $t_s-t^+=-0.25\ \mu$s (green) and $t_s-t^+=0.25\ \mu$s (blue). (b) Probability $P_{49}$ to detect the atom in $|49C\rangle$ as a function of $t_s-t^+$. The left scale converts the probability $P_{49}$ into values of $f(t^+)-f(t^-)$ using Eq. (\ref{eqn-cos}). The black dots are the experimental data, the red line corresponds to the expected values of $P_{49}$. The inset is a zoom on the transient around $t_s-t^+=0$. The vertical red, green and blue dotted lines correspond to the values of $t_s-t^+$ used in panel (a). (c) Frequency response of the electrometer. Relative sensitivity, $G(\nu)/G_0$, as a function of the signal frequency $\nu$. The dots are experimental, the line results from a simulation of the experiment.} 
\label{fig-2}
\end{figure}

We record the oscillations of $P_{49}$ with $\varphi$ and measure their amplitude, $\tilde P_{49}(\nu)$, as a function of $\nu$. We thus get the frequency response $G(\nu)=\tilde P_{49}(\nu)/ \tilde f(\nu)$.  Fig. 2(c) presents $G(\nu)/G_0$ as a function of $\nu$ (dots). The solid line presents the theoretical predictions of a model taking into account the explicit dipole time dependency during the rf pulses. It is in excellent agreement with the measured data, except around 5~MHz, due to uncertainties on the frequency response of the transmission lines, which feed the capacitor defining $\mathbf{F}$~\cite{supplementary}. Globally, the electrometer behaves as a low-pass filter with a 3~dB cut-off frequency of 5~MHz, i.e. a response time of 200~ns.  The bandwidth is only limited by the total duration of the rf pulses, as intuitively expected.

We now  determine the correlation function of a stochastic field $f(t)$ (with a zero average value). The accumulated phase, $\Phi_{tot}\simeq\alpha[f(t^+)-f(t^-)]$, is now random. The Ramsey fringes, averaged over many realizations of the experiment, have a reduced contrast, $C_0C_r$, with $C_r<1$. For a small signal amplitude ($\Phi_{tot}\ll\pi$), we get at the second order in this amplitude~\cite{supplementary}
\begin{equation}
C_r\approx 1-\left (\alpha \sigma_f \right)^2\left[1 - G^1_f(t^--t^+)\right]\ ,
\label{eq:corr}
\end{equation}
where   $\sigma_f ^2= \langle f(t^+) ^2 \rangle$ is the field variance (the brackets denote an ensemble average of noise realizations) and where $G^1_f(\tau)=(\langle f(t)f(t-\tau)\rangle)/\langle f(t)^2 \rangle$
is the first-order time-correlation function of $f$. In this regime, the Ramsey fringes contrast directly measures the field statistical properties. We still get an explicit expression of $C_r$ in the case of a large signal amplitude, provided it has a Gaussian distribution
\begin{equation}
C_r=\exp\left[-\alpha^2 \sigma_f ^2\left(1 - G^1_f(t^--t^+)\right)\right]\ .
\label{eq:correx}
\end{equation}

We apply on the atoms a numerically generated Gaussian stochastic signal with a tunable amplitude and an exponential correlation with a characteristic time $\tau_c=1.5\ \mu$s. We set $t^--t^+=9\ \mu$s, so that $f(t^+)$ and $f(t^-)$ are uncorrelated ($G^1_f\approx 0$). We plot on Fig. 3 the fringe contrast reduction, $C_r$, as a function of $\sigma_f $. The dotted parabola corresponds to the predictions of Eq.~(\ref{eq:corr}) and the solid line to the exact expression (\ref{eq:correx}). Its agreement with the measured values is excellent.

\begin{figure}

\includegraphics[width=.9\linewidth]{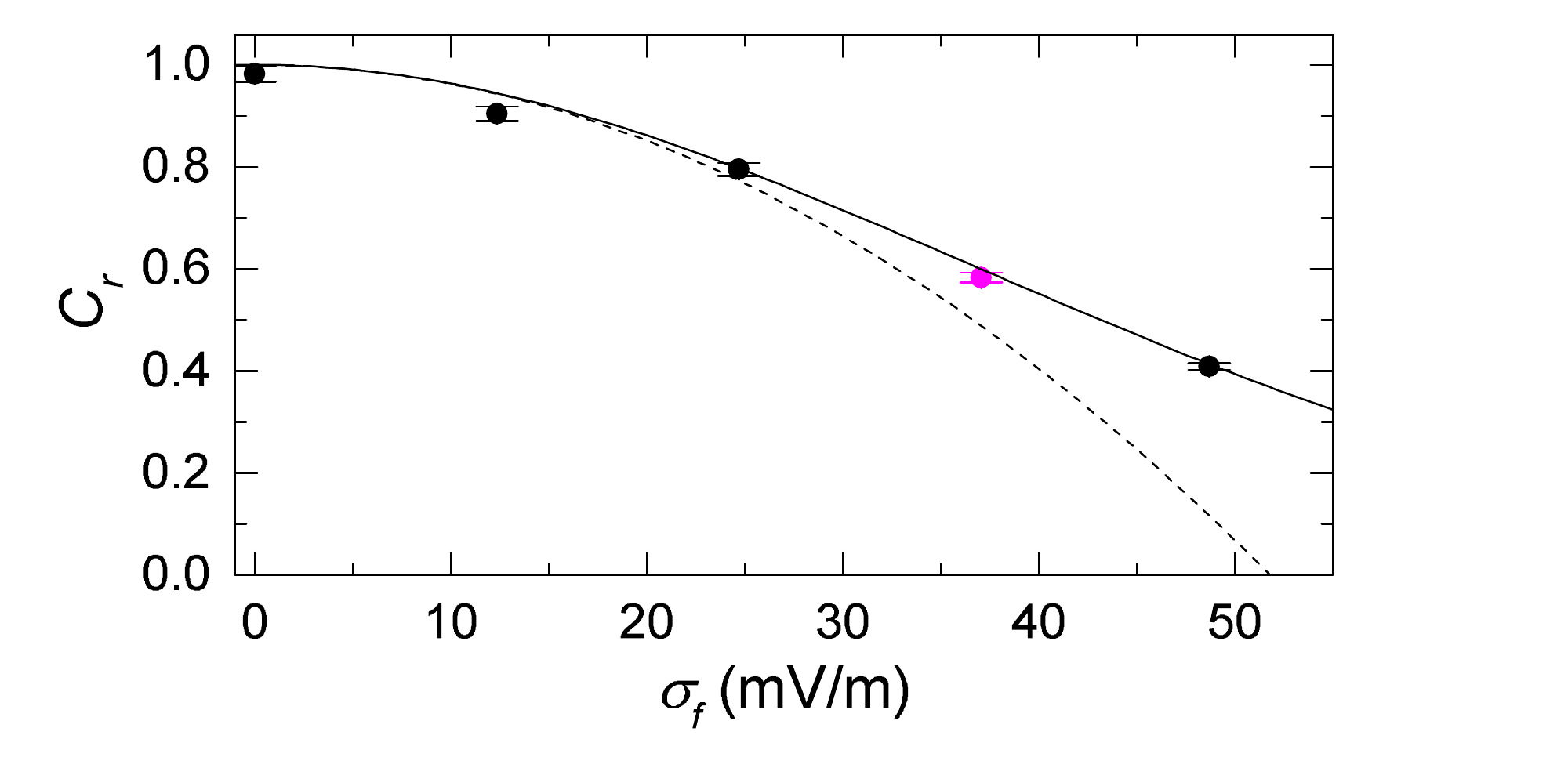}
\caption{Contrast reduction, $C_r$, as a function the stochastic field amplitude, $\sigma_f$. The solid points correspond to the contrast of the fringes in the presence of noise normalized to $C_0$. The dashed line correspond to the predictions of Eq. (\ref{eq:corr}), the solid line to the exact expression [Eq. (\ref{eq:correx})].  The magenta point corresponds to the noise amplitude used in Fig. 4. }
\label{fig-3}
\end{figure}

We show in   Fig.~4(a) the contrast, $C_0C_r$, as a function of $\tau=(t^--t^+)$.  The black dots correspond to a measurement with no applied noise ($C_r=1$). As required, the contrast, $C_0$, of the fringes does not depend on $\tau$ (within the error bars).  We then apply three Gaussian stochastic signals with the same r.m.s. amplitude $\sigma_f=37$~mV/m (magenta dot on Fig. 3). The red dots correspond to $\tau_c=1.5\ \mu$s. The red solid line corresponds to the predictions of  Eq.~(\ref{eq:correx}), using the known correlation function $G^1_f(\tau)$. The green dots and green lines depict respectively the experimental data and the theoretical predictions for $\tau_c=0.5\ \mu$s. Finally, the blue dots present the contrast obtained for a signal with a short correlation time ($G^1_f(\tau)= 0$ for $\tau>200$~ns~\cite{supplementary}). The agreement between expectations and measurements shows that the electrometer faithfully measures the field correlation function down to a time scale limited by the duration of the rf pulses.

\begin{figure}

\includegraphics[width=.9\linewidth]{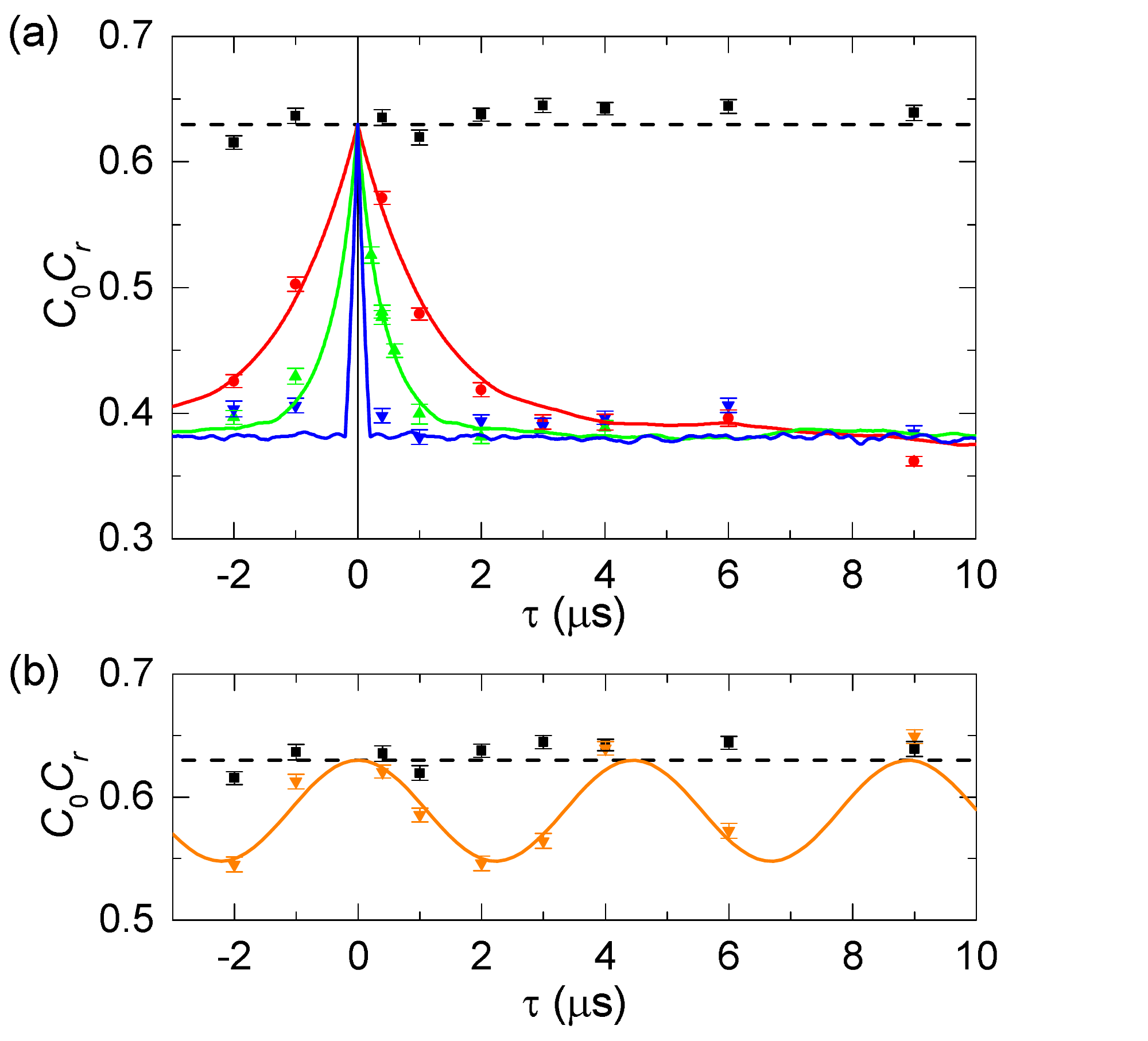}

\caption{ (a) Contrast, $C_0C_r$, as a function of $\tau=(t^--t^+)$. Black dots: no applied noise. The horizontal dotted line corresponds to $C_0=0.63$. Red and green dots: experimental data for a stochastic signal with a noise amplitude $\sigma_f=37$~mV/m and correlation times $\tau_c=1.5\ \mu$s and $\ 0.5\ \mu$s respectively. Blue dots : experimental data for a stochastic signal with no correlations for $\tau>200$~ns. Solid red, green and blue lines: corresponding theoretical predictions.  (b) Black dots and dotted black line: same as in panel (a). Orange dots and solid orange line: experimental data and theoretical predictions for a sine signal asynchronous with the experiment.
}
\label{fig-4}
\end{figure}

The field correlation function can also be used to measure a deterministic signal that is not synchronized with the experiment, a frequent situation in electrometer applications. We apply a sinusoidal field, $f(t)$, at a frequency $\nu=225$~kHz, with a small amplitude 18.8~mV/m. The experimental sequence is triggered randomly with respect to this signal. A field measurement performed by averaging many realizations of the sequence leads thus to a null result. The dots in Fig.~4(b) present the Ramsey fringes contrast as a function of $\tau$. The sinusoidal variation reflects the sinusoidal correlation of the signal. The solid line corresponds to the predictions of Eq. (\ref{eq:corr}), in excellent agreement with the data over the whole range of $\tau$ values.

We have demonstrated a quantum-enabled differential electrometer relying on the  full breadth of the Rydberg Stark manifold, using in sequence atomic states with large positive or negative dipoles. It measures the correlation function of small electric fields with a high bandwidth and a high sensitivity. The single-atom sensitivity is $83.2\pm1.4$~mV/m, well below the SQL. It corresponds to the field of a single charge at a 130 $\mu$m distance.  The time resolution reaches 200~ns, set by the finite duration of the rf pulses. 

The method relies on arbitrary manipulations of the sign and of the amplitude of the electric dipole. It could be used to measure multi-time correlation functions of the field by alternating between positive and negative dipoles. More generally, it could be used to adapt the frequency response of the electrometer to the measured signal, while reducing the adverse influence of noise at other frequencies.

The Rydberg electrometer could benefit from the high-$\ell$ Rydberg states laser trapping, through the ponderomotive potential~\cite{ENS_CIRCSIM18}. This opens the way to the development of non-invasive local probes to study the charge dynamics inside mesoscopic devices, or to characterize electric field noise above a surface with a high spatial resolution ~\cite{ION_BLATTRMP15}.

\begin{acknowledgements}
We acknowledge financial support by the European Union under the Research and Innovation action project
    ``RYSQ'' (640378) and by the Agence Nationale de la Recherche under the project  ``SNOCAR'' (167754).
\end{acknowledgements}

\renewcommand{\theequation}{S\arabic{equation}}
\renewcommand{\thefigure}{S\arabic{figure}}
\setcounter{figure}{0}

\clearpage

\section{Supplementary : Quantum-enabled electrometer measures field transients and correlation function}

We detail here the effect of the radiofrequency pulses on the $|49C\rangle$ state, the calibration of the sensitivity, the calibration of the frequency response of the driving lines of the cryostat and the generation of the electric field noise that we apply on the atoms. We also present the calculation of the fringes contrast for a stochastic signal and define the Standard Quantum Limit adapted to this measurement scheme.

\subsection{A. Atomic dynamics in the $n=49$ manifold}

At the electric field $F_0=2.345$~V/cm, the Stark frequency in the $n=49$ manifold is $\omega_{49}/2\pi=221$~MHz. Therefore, the 230 MHz rf field is not resonant with the transitions between Stark levels in this manifold. The dynamics of the atom initially in the $|49C\rangle$ state driven by the $\sigma^\pm$ rf pulses is the off-resonant Rabi oscillation of an angular momentum. It rotates around a direction close to the vertical axis of its Bloch sphere at a frequency $\tilde{\Omega} = \sqrt{\Omega_{49}^2+\delta^2}$ , where $\delta = \omega_{49}-\omega_{rf}$ and $\Omega_{49}$ is proportional to the amplitude of the rf field. During the rf pulse, the atom initially in $|49C\rangle$ will transiently populate the neighboring elliptic states before returning in $|49C\rangle$ after a time $t_{rf} = t_{2\pi} = 2\pi/\tilde{\Omega}$. We choose the duration $t_{rf}$ of the rf pulses to be precisely $t_{2\pi}$. 

Because of the second order Stark shift, the spacing of the levels of the upper diagonal [in orange on Fig.~1(a) in the main text] is slightly smaller than the spacing of the levels of the lower diagonal [in blue on Fig. 1(a)] near the circular state. As a result, the effective value of the detuning $\delta$ is different for  $\sigma^+$ and $\sigma^-$ polarizations, explaining why we use different values for the duration of the rf pulses at $t+$ and $t^-$, $t_{rf}^+$ and $ t_{rf}^-$. Experimentally, we find $t_{rf}^+=102$~ns and $t_{rf}^-=97$~ns.   

\subsection{B. Calibration of $\alpha_\pm$}

In order to calibrate $\alpha_+$ and $\alpha_-$, we use three sequences. In the first sequence, (a), we only apply the two $\pi/2$ microwave pulses resonant with the $51C\rightarrow49C$ transition at $t=9\ \mu$s and $t=35\ \mu$s. In the second sequence, (b),  we apply, in addition to the microwave pulses, the two $\sigma^+$ polarized rf pulses at $t=t^+$. In the third sequence, (c), we apply instead the two $\sigma^-$ polarized rf pulses at $t=t^-$. We record the Ramsey fringes corresponding to the sequences (a), (b) and (c) for  two different values of the electric field, $F_\pm = F_0\pm \delta \! f_0/2$, with $ \delta \! f_0=75.2$~mV/m, and measure for each sequence the phase shift ($\Delta\Phi_{a,b,c}$) induced on the fringes by the change of the electric field by $\delta \! f_0$. The difference between $\Delta\Phi_{b}$ ($\Delta\Phi_{c}$) and $\Delta\Phi_{a}$ is a direct measurement of the phase shift $\alpha_+\delta \! f_0$ ($\alpha_-\delta \! f_0$) accumulated during the $\sigma^+$ ($\sigma^-$) rf pulses.

\subsection{C. Calibration of the frequency response of the cryostat transmission lines}

In order to measure the frequency response of the electrometer, we apply on the atom an electric field $f(t)$, null around $t^+$  and sinusoidal at a frequency $\nu$ near $t=t^-$ [Fig.~S1(a)]. To create this field, we generate, starting at time $t_s>t^+$, a sinusoidal voltage $V(t)$ with amplitude $V_0$ that we add, through a 100 k$\Omega$ resistor, to the electric line connected to one of the plane electrodes of the capacitor defining the electric field $\mathbf{F}$. To determine the amplitude of the electric field seen by the atom, we measure on an oscilloscope the amplitude of the voltage after the 100 k$\Omega$ resistor. From this measurement and from the electrodes geometry, we  deduce the amplitude $\tilde f(\nu) = \lambda(\nu) V_0$ of the sinusoidal electric field at the position of the atom. Due to the electrical resonances of the transmission line, the proportionality factor $\lambda(\nu)$ between this amplitude and the applied voltage $V_0$ depends on the frequency $\nu$. Fig. S1(b) presents the experimental values of $\lambda(\nu)/\lambda(0)$, where $\lambda(0)$ is measured for a DC signal applied on the resistor. A sharp resonance of the transmission line around 8.3~MHz is conspicuous.

\begin{figure}
\includegraphics[width=.9\linewidth]{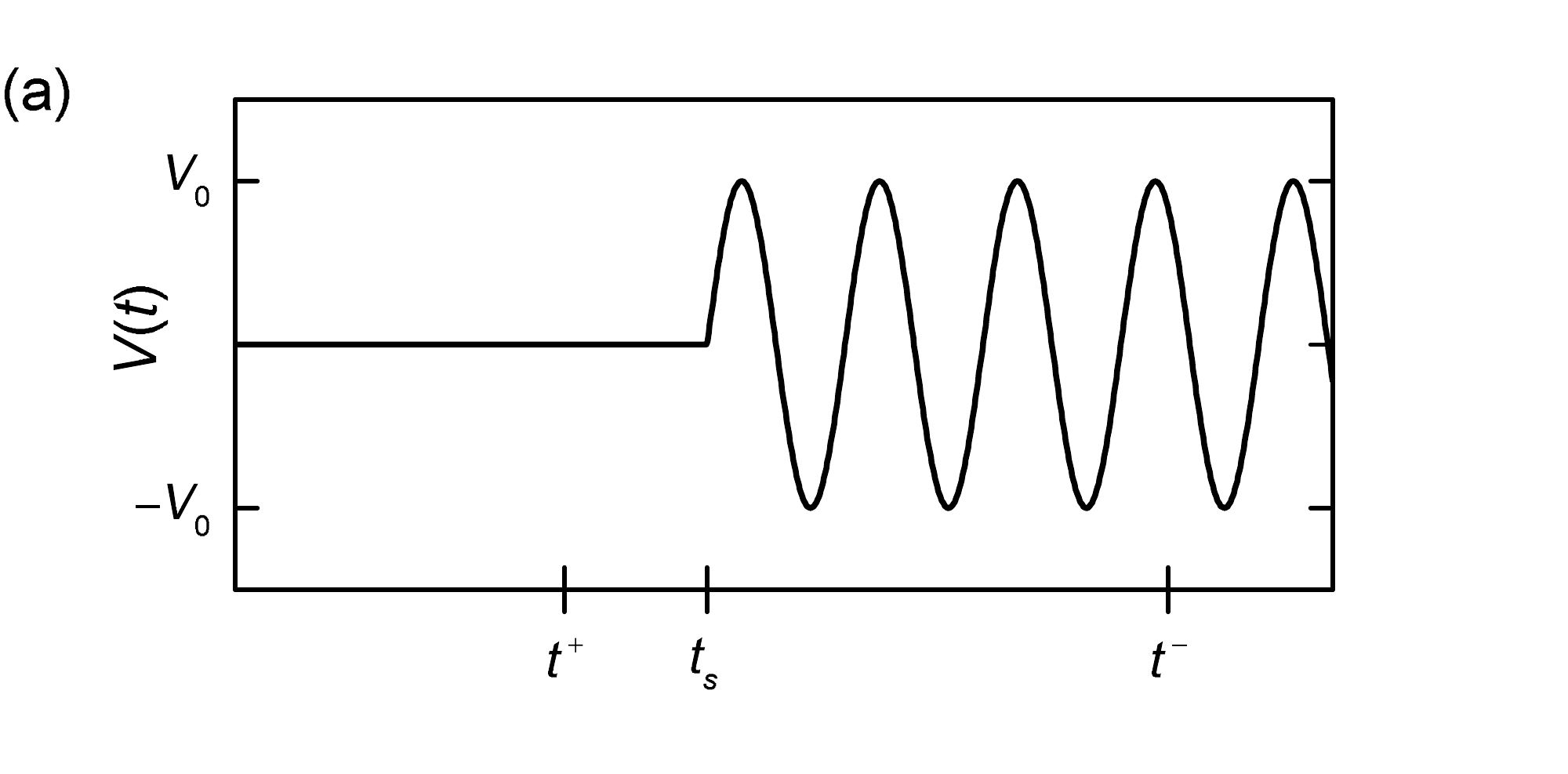}\\
\vspace{-0.5cm}
\includegraphics[width=.9\linewidth]{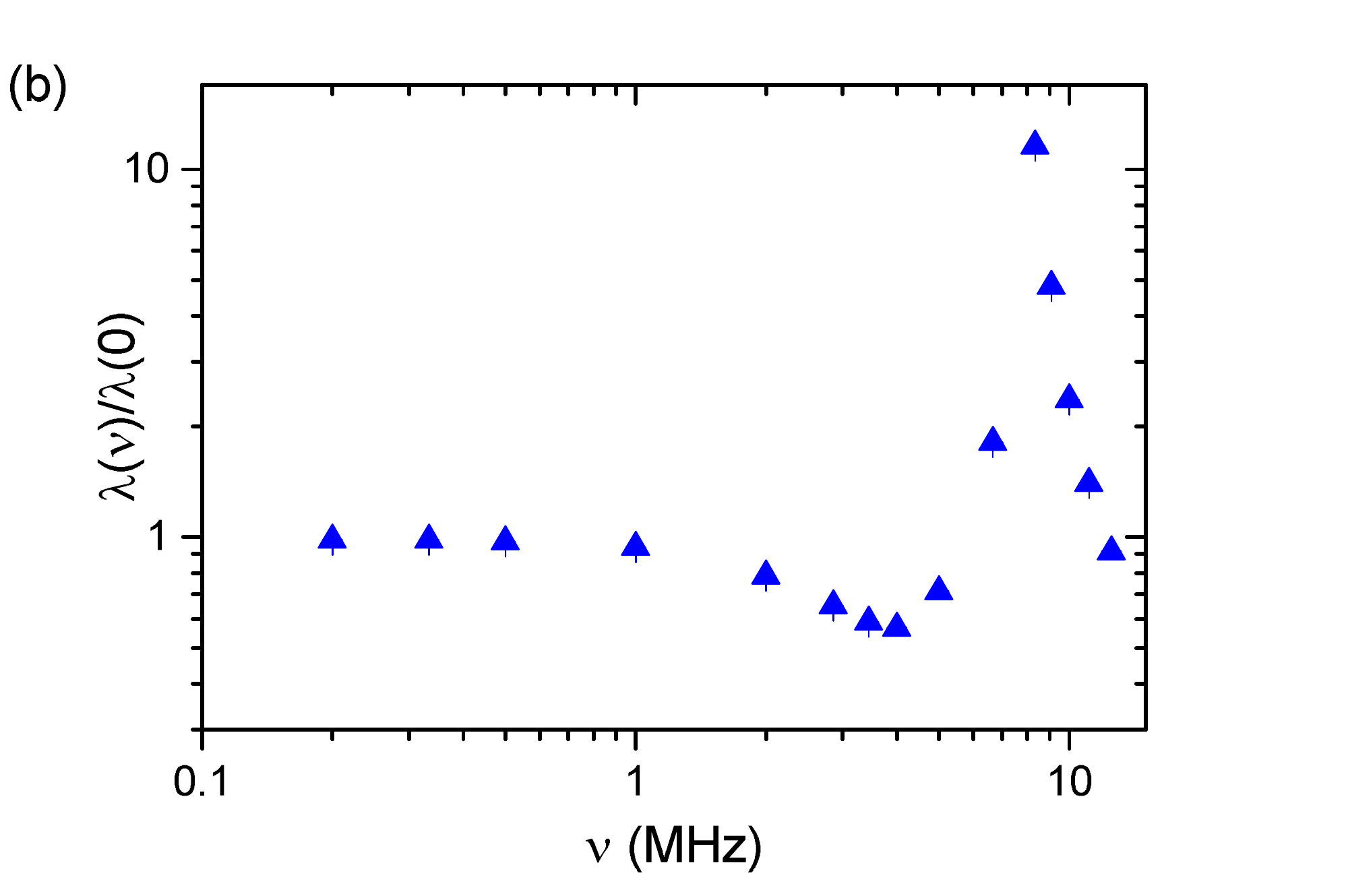}\\
\vspace{-0.3cm}
\includegraphics[width=.9\linewidth]{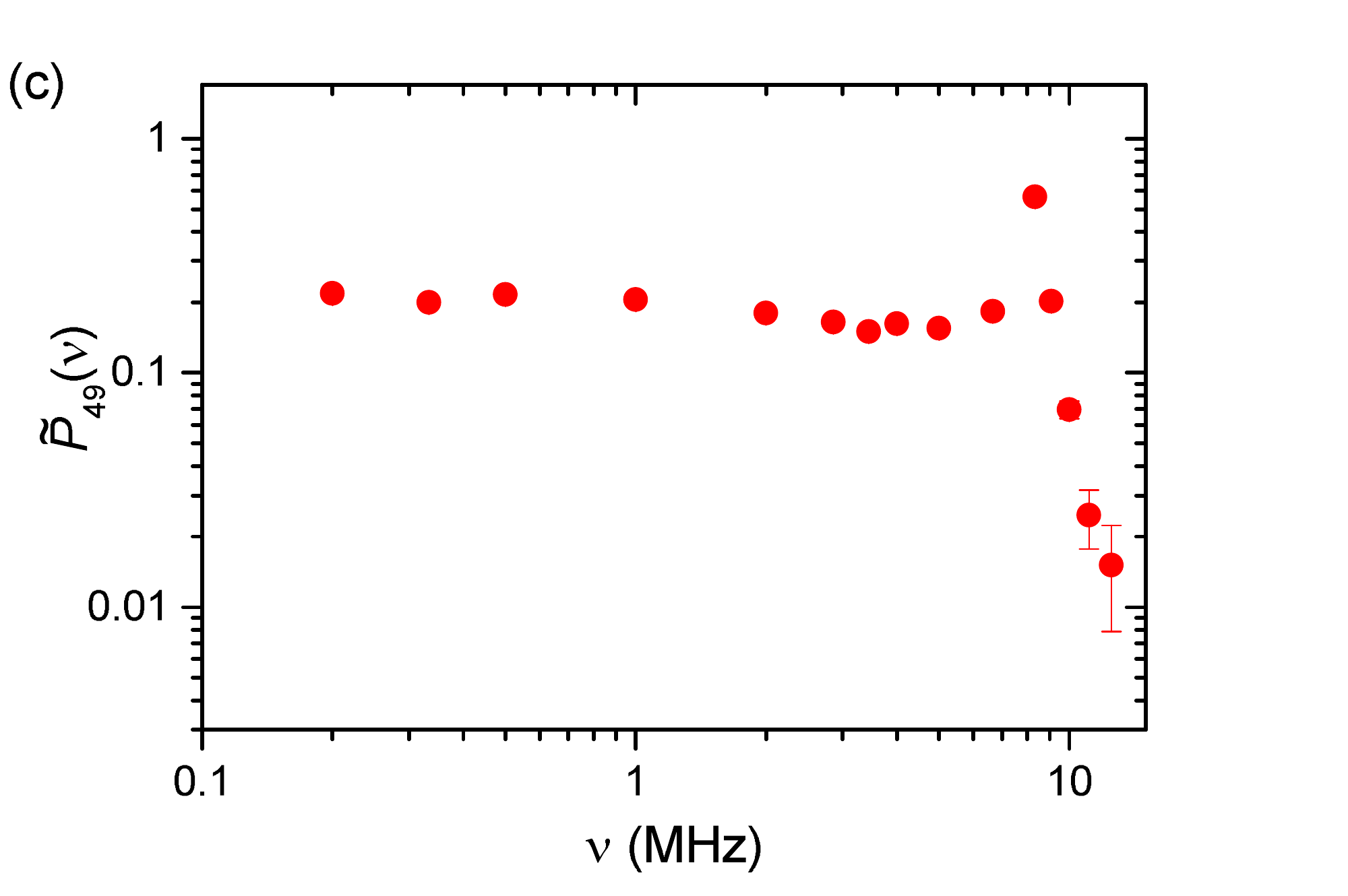}
\caption{ (a) Signal applied to calibrate the frequency response. (b) Transmission line response,  $\lambda(\nu)/\lambda(0)$, as a function of the signal frequency $\nu$. (c)  Amplitude $\tilde P_{49}(\nu)$ of the sinusoidal variation of the probability for detecting the atom in $|49C\rangle$ as a function of the signal frequency $\nu$.}
\label{fig-1}
\end{figure}

We then record the probability $P_{49}(t_s)$ to detect the atom in $|49C\rangle$ as a function of $t_s$. Changing $t_s$ varies the phase of the signal $f$ at $t=t^-$. $P_{49}(t_s)$ thus varies sinusoidally with $t_s$. We extract the amplitude $\tilde P_{49}(\nu)$ from a numerical fit of the experimental data [Fig. S1(c)]. Finally, we deduce the electrometer frequency response $G(\nu)=\tilde P_{49}(\nu)/ \tilde f(\nu)$, plotted on Fig.~2(c) in the main paper.
Note that at the peak of the 8.3~MHz resonance, $\tilde P_{49}(\nu)$ reaches 0.57. Deviations from the linear regime are expected. We thus have excluded this point from the data in Fig.~2(c). 

\subsection{D. Generation of the stochastic electric field }

In order to create the stochastic signal, $f(t)$, we use an arbitrary waveform generator (AWG), which outputs a signal equal to a tunable voltage $V_0$ multiplied by a programmable sequence $-1\le S(i) \le 1$, extracted from its memory at a 100 MS/s clock rate.  The AWG  is connected to the plane electrode through the 100 k$\Omega$ resistor. We first generate 131072 random values $y_0(i)$ with a Gaussian distribution. We apply a numerical low-pass filter to create the sequence 
$$y_{j_0}(i) = \sum_{j=0}^{131071} e^{-j/j_0} y_0(i-j)$$ 
with a correlation constant $j_0=150$. We divide $y_{j_0}(i)$ by the largest value $y_{max}$ of $|y_{j_0}(i)|$ to create the final sequence 
$S_{j_0}(i) = {y_{j_0}(i)}/{y_{max}} $, with a zero average and a variance $ \sigma_{S_{j_0}}=0.259$. We load $S_{j_0}$ in the AWG memory and program $V_0$  on the AWG interface. 

We play the sequence in a loop, generating a voltage $V_0 S_{j_0}([t (ns)/10]) $, with a time-correlation function that decays exponentially with a $\tau_c=1.5\ \mu$s time constant. Most of the spectral components of the signal are at frequencies $\nu\le1/(2\pi\tau_c)$, for which $\lambda(\nu)\approx \lambda(0)$. The stochastic electric field signal applied on the atom thus reads $f(t)=\lambda(0)V_0 S_{j_0}(t) $. The total duration of the sequence, 1.31072 ms, is incommensurable with the duration of the experimental sequence (311 $\mu$s). As we repeat many times the experiment, the RF pulses randomly samples $f(t)$ with, on the average, $\langle f(t) \rangle \approx 0$ and $\sigma_f =  \lambda(0) V_0 \sigma_{S_{j_0}} $.

The stochastic field with a correlation time $\tau_c=0.5\ \mu$s is generated using the same method, with a new set of random values $y^\prime_0(i)$ and $j_0=50$. Finally, the ``white noise" [blue dots on Fig.~4(a) in the main text] is generated using 131072 random values $y^{\prime\prime}_0(i)$ that are averaged on a 20 points sliding window
$$y_{white}(i) = \frac 1 {20} \sum_{j=0}^{19} y^{\prime\prime}_0(i-j)$$ 
before being normalized to the maximum value of $y_{white}(i)$ and sent to the AWG. As a result, there is no correlation between two values of $y_{white}$ separated by more than 20 points (or 200 ns). The 20-points averaging removes spectral components around $\nu\approx  8.3$ MHz, where we observed a sharp resonance in $\lambda(\nu)$ [Fig.S1(b)]. This allows us to  estimate $\sigma_{f}$ using again the approximation $\sigma_{f} =  \lambda(0) V_0 \sigma_{S_{white}} $.

\subsection{E. Calculation of $C_r$}

In the presence of a stochastic electric field, $f(t)$, the Ramsey fringes are given by
\begin{equation}
	P_{49}=\left\langle P_0+\frac{C_0}{2}\cos(\Phi_{tot}-\varphi_{mw})\right\rangle\ ,
	\label{eqn-av}
\end{equation}
where $\langle \cdot \rangle$ denote the averaging over many realizations of $f(t)$ and $\Phi_{tot}=\alpha[f(t^+)-f(t^-)]$. It is convenient to rewrite Eq. (\ref{eqn-av}) as
$$
	P_{49}=  \Re \left\{ P_0+\frac{C_0}{2} \left\langle  e^{i\alpha[f(t^+)-f(t^-)]} \right\rangle e^{-i\varphi_{mw}}\right\}\ .
$$
Finally, we find
$$
	P_{49}= P_0+\frac{C_0}{2} C_r \cos(\Phi-\varphi_{mw})\ ,
$$
where $C_r$ and $\Phi$ are respectively the modulus and the phase of $ \langle e^{i\alpha[f(t^+)-f(t^-)]} \rangle$. For small values of $[f(t^+)-f(t^-)]$, and since $\langle [f(t^+)-f(t^-)]\rangle = 0$, we can use the Taylor expansion
$$ \langle e^{i\alpha[f(t^+)-f(t^-)]} \rangle \approx 1 - \frac1 2 \alpha^2\langle [f(t^+)-f(t^-)]^2\rangle$$
We find that $\Phi=0$. We can expand $[f(t^+)-f(t^-)]^2$ to get the reduced contrast $C_r$
$$ C_r \approx 1 - \frac{\alpha^2} 2 \left [ \langle f(t^+)^2\rangle +\langle f(t^-)^2\rangle - 2\langle f(t^+)f(t^-)\rangle \right]$$
We then obtain Eq. (2) in the main text using $ \langle f(t^+)^2\rangle =\langle f(t^-)^2\rangle =\sigma_f^2$.

If $f(t^+)$ and $f(t^-)$ are two Gaussian random variables, $\delta \! f =f(t^+)- f(t^-)$ is also a gaussian random variable with variance $\sigma_{\delta \! f}^2=2\sigma_f^2 \left [ 1 - G^1_f(t^--t^+) \right]$. The reduced contrast $C_r$ can then be expressed as
$$ C_r= \left | \langle e^{i\alpha[f(t^+)-f(t^-)]} \rangle \right | = \frac{1}{\sqrt{2\pi}\sigma_{\delta \! f}} \left |\int_{-\infty}^{\infty} e^{i\alpha u} e^{-\frac{u^2}{2\sigma_{\delta \! f}^2}} \mbox d u\right |$$
so that, finally, $$C_r = \exp\left[-\alpha^2 \sigma_f ^2\left(1 - G^1_f(t^--t^+)\right)\right]\ .$$

\subsection{F. Standard quantum limit}

The principle of the differential electrometer is to infer the value of $\delta f =f(t^+)- f(t^-)$ from the measurement of the state of the atom. The uncertainty on $\delta f$ obtained from the detection of one atom is given by 
$$\sigma_{\delta \! f}^1=\left| \frac{dP_{49}}{d\,\delta \! f} \right|^{-1}\sigma_{P_{49}} \ ,$$
where $\sigma_{P_{49}}= \sqrt{P_{49}(1-P_{49})} $ is the standard deviation of a single atomic detection. When averaging the results of $k$ atomic detections, the uncertainty becomes $\sigma_{ \delta \! f}^k = \sigma_{\delta \! f}^1/\sqrt k$. For small values of $\delta \! f$, $P_{49}\approx 1/2$, therefore $\sigma_{P_{49}}\approx 1/2 $. Using $|dP_{49}/d\,\delta \! f| = G_0=\alpha C_0 / 2$, we find 
$$\sigma_{\delta \! f}^1=\frac 1 {\alpha C_0} = 82.3 \pm 1.4 \ \mbox{mV/m}.$$

A naive classical method to reach the Standard Quantum Limit would correspond to two separate measurements of $f(t^+)$ and $f(t^-)$ performed at the SQL by manipulating independently spin coherent states of the spins $J_1$ and $J_2$. Each measurement, with a duration $T$, achieves a SQL sensitivity~\cite{Facon2016}
$$
\sigma_{SQL}^1=\frac{1}{T\sqrt{n-1}}\left|\frac{\partial\omega}{\partial F}  \right|^{-1}\ .
\label{eqn-sql}
$$
These separate measurements are combined to compute $\delta f$ with a single atom uncertainty $\sqrt 2\sigma^1_{SQL}$

At least in the case $t^--t^+=T$, we can imagine a more efficient strategy using spin coherent states of the spin $J_1$ alone. Starting from the north pole of its Bloch sphere, we perform a $\pi/2$ rotation bringing it in the equatorial plane along the $x$ axis. During a first interrogation time $T$ centered at $t^+$, the spin accumulates a phase shift $T f(t^+)\partial\omega/\partial F$. We then perform a spin echo with a $\pi$ pulse around the $x$ axis and let the spin rotate in the field for another time interval $T$ centered at $t^-$. The final spin phase rotation is thus $T \delta f\partial\omega/\partial F$, which can be measured with a $1/\sqrt {2J_1}$ uncertainty~\cite{Facon2016}. The sensitivity of a classical measurement of $\delta f$ can thus reach $\sigma_{SQL}^1$, $\sqrt 2$ times lower than that of the naive differential measurement.

For $T = 206$ ns, we get $\sigma_{SQL}^1= 111.5$ mV/m. The sensitivity of our electrometer is therefore $2.64 \pm 0.15$ dB below the SQL. It is limited by the finite fringes contrast, but also by the finite duration of the rf pulses : even though the electric dipole, $D(t)$, reach the maximum value of $D_{max}\sim (3/2)n^2 a_0 e$ at $t=t^\pm$, the average value of $D(t)$ over the duration $T$ of the measurement is $\langle D(t)\rangle \approx D_{max}/2$.


\begin{thebibliography}{27}
\expandafter\ifx\csname natexlab\endcsname\relax\def\natexlab#1{#1}\fi
\expandafter\ifx\csname bibnamefont\endcsname\relax
  \def\bibnamefont#1{#1}\fi
\expandafter\ifx\csname bibfnamefont\endcsname\relax
  \def\bibfnamefont#1{#1}\fi
\expandafter\ifx\csname citenamefont\endcsname\relax
  \def\citenamefont#1{#1}\fi
\expandafter\ifx\csname url\endcsname\relax
  \def\url#1{\texttt{#1}}\fi
\expandafter\ifx\csname urlprefix\endcsname\relax\def\urlprefix{URL }\fi
\providecommand{\bibinfo}[2]{#2}
\providecommand{\eprint}[2][]{\url{#2}}

\bibitem[{\citenamefont{Lu et~al.}(2003)\citenamefont{Lu, Ji, Pfeiffer, West,
  and Rimberg}}]{Lu2003}
\bibinfo{author}{\bibfnamefont{W.}~\bibnamefont{Lu}},
  \bibinfo{author}{\bibfnamefont{Z.}~\bibnamefont{Ji}},
  \bibinfo{author}{\bibfnamefont{L.}~\bibnamefont{Pfeiffer}},
  \bibinfo{author}{\bibfnamefont{K.~W.} \bibnamefont{West}}, \bibnamefont{and}
  \bibinfo{author}{\bibfnamefont{A.~J.} \bibnamefont{Rimberg}},
  \bibinfo{journal}{Nature} \textbf{\bibinfo{volume}{423}},
  \bibinfo{pages}{422} (\bibinfo{year}{2003}).

\bibitem[{\citenamefont{Bylander et~al.}(2005)\citenamefont{Bylander, Duty, and
  Delsing}}]{Bylander2005}
\bibinfo{author}{\bibfnamefont{J.}~\bibnamefont{Bylander}},
  \bibinfo{author}{\bibfnamefont{T.}~\bibnamefont{Duty}}, \bibnamefont{and}
  \bibinfo{author}{\bibfnamefont{P.}~\bibnamefont{Delsing}},
  \bibinfo{journal}{Nature} \textbf{\bibinfo{volume}{434}},
  \bibinfo{pages}{361} (\bibinfo{year}{2005}).

\bibitem[{\citenamefont{Fujisawa et~al.}(2006)\citenamefont{Fujisawa, Hayashi,
  Tomita, and Hirayama}}]{Fujisawa2006}
\bibinfo{author}{\bibfnamefont{T.}~\bibnamefont{Fujisawa}},
  \bibinfo{author}{\bibfnamefont{T.}~\bibnamefont{Hayashi}},
  \bibinfo{author}{\bibfnamefont{R.}~\bibnamefont{Tomita}}, \bibnamefont{and}
  \bibinfo{author}{\bibfnamefont{Y.}~\bibnamefont{Hirayama}},
  \bibinfo{journal}{Science} \textbf{\bibinfo{volume}{312}},
  \bibinfo{pages}{1634} (\bibinfo{year}{2006}).

\bibitem[{\citenamefont{Gustavsson et~al.}(2006)\citenamefont{Gustavsson,
  Leturcq, Simovi{\v{c}}, Schleser, Ihn, Studerus, Ensslin, Driscoll, and
  Gossard}}]{Gustavsson2006}
\bibinfo{author}{\bibfnamefont{S.}~\bibnamefont{Gustavsson}},
  \bibinfo{author}{\bibfnamefont{R.}~\bibnamefont{Leturcq}},
  \bibinfo{author}{\bibfnamefont{B.}~\bibnamefont{Simovi{\v{c}}}},
  \bibinfo{author}{\bibfnamefont{R.}~\bibnamefont{Schleser}},
  \bibinfo{author}{\bibfnamefont{T.}~\bibnamefont{Ihn}},
  \bibinfo{author}{\bibfnamefont{P.}~\bibnamefont{Studerus}},
  \bibinfo{author}{\bibfnamefont{K.}~\bibnamefont{Ensslin}},
  \bibinfo{author}{\bibfnamefont{D.}~\bibnamefont{Driscoll}}, \bibnamefont{and}
  \bibinfo{author}{\bibfnamefont{A.}~\bibnamefont{Gossard}},
  \bibinfo{journal}{Physical review letters} \textbf{\bibinfo{volume}{96}},
  \bibinfo{pages}{076605} (\bibinfo{year}{2006}).

\bibitem[{\citenamefont{Blanter and B{\"u}ttiker}(2000)}]{Blanter2000}
\bibinfo{author}{\bibfnamefont{Y.}~\bibnamefont{Blanter}} \bibnamefont{and}
  \bibinfo{author}{\bibfnamefont{M.}~\bibnamefont{B{\"u}ttiker}},
  \bibinfo{journal}{Physics Reports} \textbf{\bibinfo{volume}{336}},
  \bibinfo{pages}{1 } (\bibinfo{year}{2000}), ISSN \bibinfo{issn}{0370-1573},
  \urlprefix\url{http://www.sciencedirect.com/science/article/pii/S0370157399001234}.

\bibitem[{\citenamefont{Jalil et~al.}(2017)\citenamefont{Jalil, Zhu, Ekanayake,
  and Ruan}}]{Jalil2017}
\bibinfo{author}{\bibfnamefont{J.}~\bibnamefont{Jalil}},
  \bibinfo{author}{\bibfnamefont{Y.}~\bibnamefont{Zhu}},
  \bibinfo{author}{\bibfnamefont{C.}~\bibnamefont{Ekanayake}},
  \bibnamefont{and} \bibinfo{author}{\bibfnamefont{Y.}~\bibnamefont{Ruan}},
  \bibinfo{journal}{Nanotechnology} \textbf{\bibinfo{volume}{28}},
  \bibinfo{pages}{142002} (\bibinfo{year}{2017}),
  \urlprefix\url{http://stacks.iop.org/0957-4484/28/i=14/a=142002}.

\bibitem[{\citenamefont{Yoo et~al.}(1997)\citenamefont{Yoo, Fulton, Hess,
  Willett, Dunkleberger, Chichester, Pfeiffer, and West}}]{Yoo1997}
\bibinfo{author}{\bibfnamefont{M.}~\bibnamefont{Yoo}},
  \bibinfo{author}{\bibfnamefont{T.}~\bibnamefont{Fulton}},
  \bibinfo{author}{\bibfnamefont{H.}~\bibnamefont{Hess}},
  \bibinfo{author}{\bibfnamefont{R.}~\bibnamefont{Willett}},
  \bibinfo{author}{\bibfnamefont{L.}~\bibnamefont{Dunkleberger}},
  \bibinfo{author}{\bibfnamefont{R.}~\bibnamefont{Chichester}},
  \bibinfo{author}{\bibfnamefont{L.}~\bibnamefont{Pfeiffer}}, \bibnamefont{and}
  \bibinfo{author}{\bibfnamefont{K.}~\bibnamefont{West}},
  \bibinfo{journal}{Science} \textbf{\bibinfo{volume}{276}},
  \bibinfo{pages}{579} (\bibinfo{year}{1997}).

\bibitem[{\citenamefont{Schoelkopf et~al.}(1998)\citenamefont{Schoelkopf,
  Wahlgren, Kozhevnikov, Delsing, and Prober}}]{Schoelkopf1998}
\bibinfo{author}{\bibfnamefont{R.}~\bibnamefont{Schoelkopf}},
  \bibinfo{author}{\bibfnamefont{P.}~\bibnamefont{Wahlgren}},
  \bibinfo{author}{\bibfnamefont{A.}~\bibnamefont{Kozhevnikov}},
  \bibinfo{author}{\bibfnamefont{P.}~\bibnamefont{Delsing}}, \bibnamefont{and}
  \bibinfo{author}{\bibfnamefont{D.}~\bibnamefont{Prober}},
  \bibinfo{journal}{Science} \textbf{\bibinfo{volume}{280}},
  \bibinfo{pages}{1238} (\bibinfo{year}{1998}).

\bibitem[{\citenamefont{Cleland and Roukes}(1998)}]{Cleland1998}
\bibinfo{author}{\bibfnamefont{A.~N.} \bibnamefont{Cleland}} \bibnamefont{and}
  \bibinfo{author}{\bibfnamefont{M.~L.} \bibnamefont{Roukes}},
  \bibinfo{journal}{Nature} \textbf{\bibinfo{volume}{392}},
  \bibinfo{pages}{160} (\bibinfo{year}{1998}).

\bibitem[{\citenamefont{Bunch et~al.}(2007)\citenamefont{Bunch, Van Der~Zande,
  Verbridge, Frank, Tanenbaum, Parpia, Craighead, and McEuen}}]{Bunch2007}
\bibinfo{author}{\bibfnamefont{J.~S.} \bibnamefont{Bunch}},
  \bibinfo{author}{\bibfnamefont{A.~M.} \bibnamefont{Van Der~Zande}},
  \bibinfo{author}{\bibfnamefont{S.~S.} \bibnamefont{Verbridge}},
  \bibinfo{author}{\bibfnamefont{I.~W.} \bibnamefont{Frank}},
  \bibinfo{author}{\bibfnamefont{D.~M.} \bibnamefont{Tanenbaum}},
  \bibinfo{author}{\bibfnamefont{J.~M.} \bibnamefont{Parpia}},
  \bibinfo{author}{\bibfnamefont{H.~G.} \bibnamefont{Craighead}},
  \bibnamefont{and} \bibinfo{author}{\bibfnamefont{P.~L.}
  \bibnamefont{McEuen}}, \bibinfo{journal}{Science}
  \textbf{\bibinfo{volume}{315}}, \bibinfo{pages}{490} (\bibinfo{year}{2007}).

\bibitem[{\citenamefont{Dolde et~al.}(2014)\citenamefont{Dolde, Doherty, Michl,
  Jakobi, Naydenov, Pezzagna, Meijer, Neumann, Jelezko, Manson
  et~al.}}]{Dolde2014}
\bibinfo{author}{\bibfnamefont{F.}~\bibnamefont{Dolde}},
  \bibinfo{author}{\bibfnamefont{M.~W.} \bibnamefont{Doherty}},
  \bibinfo{author}{\bibfnamefont{J.}~\bibnamefont{Michl}},
  \bibinfo{author}{\bibfnamefont{I.}~\bibnamefont{Jakobi}},
  \bibinfo{author}{\bibfnamefont{B.}~\bibnamefont{Naydenov}},
  \bibinfo{author}{\bibfnamefont{S.}~\bibnamefont{Pezzagna}},
  \bibinfo{author}{\bibfnamefont{J.}~\bibnamefont{Meijer}},
  \bibinfo{author}{\bibfnamefont{P.}~\bibnamefont{Neumann}},
  \bibinfo{author}{\bibfnamefont{F.}~\bibnamefont{Jelezko}},
  \bibinfo{author}{\bibfnamefont{N.~B.} \bibnamefont{Manson}},
  \bibnamefont{et~al.}, \bibinfo{journal}{Phys. Rev. Lett.}
  \textbf{\bibinfo{volume}{112}}, \bibinfo{pages}{097603}
  (\bibinfo{year}{2014}),
  \urlprefix\url{https://link.aps.org/doi/10.1103/PhysRevLett.112.097603}.

\bibitem[{\citenamefont{Vamivakas et~al.}(2011)\citenamefont{Vamivakas, Zhao,
  F{\"a}lt, Badolato, Taylor, and Atat{\"u}re}}]{Vamivakas2011}
\bibinfo{author}{\bibfnamefont{A.}~\bibnamefont{Vamivakas}},
  \bibinfo{author}{\bibfnamefont{Y.}~\bibnamefont{Zhao}},
  \bibinfo{author}{\bibfnamefont{S.}~\bibnamefont{F{\"a}lt}},
  \bibinfo{author}{\bibfnamefont{A.}~\bibnamefont{Badolato}},
  \bibinfo{author}{\bibfnamefont{J.}~\bibnamefont{Taylor}}, \bibnamefont{and}
  \bibinfo{author}{\bibfnamefont{M.}~\bibnamefont{Atat{\"u}re}},
  \bibinfo{journal}{Phys. Rev. Lett.} \textbf{\bibinfo{volume}{107}},
  \bibinfo{pages}{166802} (\bibinfo{year}{2011}).

\bibitem[{\citenamefont{Houel et~al.}(2012)\citenamefont{Houel, Kuhlmann,
  Greuter, Xue, Poggio, Gerardot, Dalgarno, Badolato, Petroff, Ludwig
  et~al.}}]{Houel2012}
\bibinfo{author}{\bibfnamefont{J.}~\bibnamefont{Houel}},
  \bibinfo{author}{\bibfnamefont{A.~V.} \bibnamefont{Kuhlmann}},
  \bibinfo{author}{\bibfnamefont{L.}~\bibnamefont{Greuter}},
  \bibinfo{author}{\bibfnamefont{F.}~\bibnamefont{Xue}},
  \bibinfo{author}{\bibfnamefont{M.}~\bibnamefont{Poggio}},
  \bibinfo{author}{\bibfnamefont{B.~D.} \bibnamefont{Gerardot}},
  \bibinfo{author}{\bibfnamefont{P.~A.} \bibnamefont{Dalgarno}},
  \bibinfo{author}{\bibfnamefont{A.}~\bibnamefont{Badolato}},
  \bibinfo{author}{\bibfnamefont{P.~M.} \bibnamefont{Petroff}},
  \bibinfo{author}{\bibfnamefont{A.}~\bibnamefont{Ludwig}},
  \bibnamefont{et~al.}, \bibinfo{journal}{Phys. Rev. Lett.}
  \textbf{\bibinfo{volume}{108}}, \bibinfo{pages}{107401}
  (\bibinfo{year}{2012}),
  \urlprefix\url{https://link.aps.org/doi/10.1103/PhysRevLett.108.107401}.

\bibitem[{\citenamefont{Arnold et~al.}(2014)\citenamefont{Arnold, Loo,
  Lema\^{\i}tre, Sagnes, Krebs, Voisin, Senellart, and Lanco}}]{Arnold2014}
\bibinfo{author}{\bibfnamefont{C.}~\bibnamefont{Arnold}},
  \bibinfo{author}{\bibfnamefont{V.}~\bibnamefont{Loo}},
  \bibinfo{author}{\bibfnamefont{A.}~\bibnamefont{Lema\^{\i}tre}},
  \bibinfo{author}{\bibfnamefont{I.}~\bibnamefont{Sagnes}},
  \bibinfo{author}{\bibfnamefont{O.}~\bibnamefont{Krebs}},
  \bibinfo{author}{\bibfnamefont{P.}~\bibnamefont{Voisin}},
  \bibinfo{author}{\bibfnamefont{P.}~\bibnamefont{Senellart}},
  \bibnamefont{and} \bibinfo{author}{\bibfnamefont{L.}~\bibnamefont{Lanco}},
  \bibinfo{journal}{Phys. Rev. X} \textbf{\bibinfo{volume}{4}},
  \bibinfo{pages}{021004} (\bibinfo{year}{2014}),
  \urlprefix\url{https://link.aps.org/doi/10.1103/PhysRevX.4.021004}.

\bibitem[{\citenamefont{Gallagher}(1994)}]{Gallagher1994}
\bibinfo{author}{\bibfnamefont{T.~F.} \bibnamefont{Gallagher}},
  \emph{\bibinfo{title}{Rydberg Atoms}}, Cambridge Monographs on Atomic,
  Molecular and Chemical Physics (\bibinfo{publisher}{Cambridge University
  Press}, \bibinfo{year}{1994}).

\bibitem[{\citenamefont{Facon et~al.}(2016)\citenamefont{Facon, Dietsche,
  Grosso, Haroche, Raimond, Brune, and Gleyzes}}]{Facon2016}
\bibinfo{author}{\bibfnamefont{A.}~\bibnamefont{Facon}},
  \bibinfo{author}{\bibfnamefont{E.-K.} \bibnamefont{Dietsche}},
  \bibinfo{author}{\bibfnamefont{D.}~\bibnamefont{Grosso}},
  \bibinfo{author}{\bibfnamefont{S.}~\bibnamefont{Haroche}},
  \bibinfo{author}{\bibfnamefont{J.-M.} \bibnamefont{Raimond}},
  \bibinfo{author}{\bibfnamefont{M.}~\bibnamefont{Brune}}, \bibnamefont{and}
  \bibinfo{author}{\bibfnamefont{S.}~\bibnamefont{Gleyzes}},
  \bibinfo{journal}{Nature} \textbf{\bibinfo{volume}{535}},
  \bibinfo{pages}{262} (\bibinfo{year}{2016}).

\bibitem[{\citenamefont{Bylander et~al.}(2011)\citenamefont{Bylander,
  Gustavsson, Yan, Yoshihara, Harrabi, Fitch, Cory, Nakamura, Tsai, and
  Oliver}}]{Bylander2011}
\bibinfo{author}{\bibfnamefont{J.}~\bibnamefont{Bylander}},
  \bibinfo{author}{\bibfnamefont{S.}~\bibnamefont{Gustavsson}},
  \bibinfo{author}{\bibfnamefont{F.}~\bibnamefont{Yan}},
  \bibinfo{author}{\bibfnamefont{F.}~\bibnamefont{Yoshihara}},
  \bibinfo{author}{\bibfnamefont{K.}~\bibnamefont{Harrabi}},
  \bibinfo{author}{\bibfnamefont{G.}~\bibnamefont{Fitch}},
  \bibinfo{author}{\bibfnamefont{D.~G.} \bibnamefont{Cory}},
  \bibinfo{author}{\bibfnamefont{Y.}~\bibnamefont{Nakamura}},
  \bibinfo{author}{\bibfnamefont{J.-S.} \bibnamefont{Tsai}}, \bibnamefont{and}
  \bibinfo{author}{\bibfnamefont{W.~D.} \bibnamefont{Oliver}},
  \bibinfo{journal}{Nature Physics} \textbf{\bibinfo{volume}{7}},
  \bibinfo{pages}{565} (\bibinfo{year}{2011}).

\bibitem[{\citenamefont{\'Alvarez and Suter}(2011)}]{Alvarez2011}
\bibinfo{author}{\bibfnamefont{G.~A.} \bibnamefont{\'Alvarez}}
  \bibnamefont{and} \bibinfo{author}{\bibfnamefont{D.}~\bibnamefont{Suter}},
  \bibinfo{journal}{Phys. Rev. Lett.} \textbf{\bibinfo{volume}{107}},
  \bibinfo{pages}{230501} (\bibinfo{year}{2011}),
  \urlprefix\url{https://link.aps.org/doi/10.1103/PhysRevLett.107.230501}.

\bibitem[{\citenamefont{Dial et~al.}(2013)\citenamefont{Dial, Shulman, Harvey,
  Bluhm, Umansky, and Yacoby}}]{Dial2013}
\bibinfo{author}{\bibfnamefont{O.}~\bibnamefont{Dial}},
  \bibinfo{author}{\bibfnamefont{M.~D.} \bibnamefont{Shulman}},
  \bibinfo{author}{\bibfnamefont{S.~P.} \bibnamefont{Harvey}},
  \bibinfo{author}{\bibfnamefont{H.}~\bibnamefont{Bluhm}},
  \bibinfo{author}{\bibfnamefont{V.}~\bibnamefont{Umansky}}, \bibnamefont{and}
  \bibinfo{author}{\bibfnamefont{A.}~\bibnamefont{Yacoby}},
  \bibinfo{journal}{Phys. Rev. Lett.} \textbf{\bibinfo{volume}{110}},
  \bibinfo{pages}{146804} (\bibinfo{year}{2013}).

\bibitem[{\citenamefont{Romach et~al.}(2015)\citenamefont{Romach, M\"uller,
  Unden, Rogers, Isoda, Itoh, Markham, Stacey, Meijer, Pezzagna
  et~al.}}]{Romach2015}
\bibinfo{author}{\bibfnamefont{Y.}~\bibnamefont{Romach}},
  \bibinfo{author}{\bibfnamefont{C.}~\bibnamefont{M\"uller}},
  \bibinfo{author}{\bibfnamefont{T.}~\bibnamefont{Unden}},
  \bibinfo{author}{\bibfnamefont{L.~J.} \bibnamefont{Rogers}},
  \bibinfo{author}{\bibfnamefont{T.}~\bibnamefont{Isoda}},
  \bibinfo{author}{\bibfnamefont{K.~M.} \bibnamefont{Itoh}},
  \bibinfo{author}{\bibfnamefont{M.}~\bibnamefont{Markham}},
  \bibinfo{author}{\bibfnamefont{A.}~\bibnamefont{Stacey}},
  \bibinfo{author}{\bibfnamefont{J.}~\bibnamefont{Meijer}},
  \bibinfo{author}{\bibfnamefont{S.}~\bibnamefont{Pezzagna}},
  \bibnamefont{et~al.}, \bibinfo{journal}{Phys. Rev. Lett.}
  \textbf{\bibinfo{volume}{114}}, \bibinfo{pages}{017601}
  (\bibinfo{year}{2015}),
  \urlprefix\url{https://link.aps.org/doi/10.1103/PhysRevLett.114.017601}.

\bibitem[{\citenamefont{Kotler et~al.}(2011)\citenamefont{Kotler, Akerman,
  Glickman, Keselman, and Ozeri}}]{Kotler2011}
\bibinfo{author}{\bibfnamefont{S.}~\bibnamefont{Kotler}},
  \bibinfo{author}{\bibfnamefont{N.}~\bibnamefont{Akerman}},
  \bibinfo{author}{\bibfnamefont{Y.}~\bibnamefont{Glickman}},
  \bibinfo{author}{\bibfnamefont{A.}~\bibnamefont{Keselman}}, \bibnamefont{and}
  \bibinfo{author}{\bibfnamefont{R.}~\bibnamefont{Ozeri}},
  \bibinfo{journal}{Nature} \textbf{\bibinfo{volume}{473}}, \bibinfo{pages}{61}
  (\bibinfo{year}{2011}).

\bibitem[{\citenamefont{Bar-Gill et~al.}(2012)\citenamefont{Bar-Gill, Pham,
  Belthangady, Le~Sage, Cappellaro, Maze, Lukin, Yacoby, and
  Walsworth}}]{BarGill2012}
\bibinfo{author}{\bibfnamefont{N.}~\bibnamefont{Bar-Gill}},
  \bibinfo{author}{\bibfnamefont{L.~M.} \bibnamefont{Pham}},
  \bibinfo{author}{\bibfnamefont{C.}~\bibnamefont{Belthangady}},
  \bibinfo{author}{\bibfnamefont{D.}~\bibnamefont{Le~Sage}},
  \bibinfo{author}{\bibfnamefont{P.}~\bibnamefont{Cappellaro}},
  \bibinfo{author}{\bibfnamefont{J.~R.} \bibnamefont{Maze}},
  \bibinfo{author}{\bibfnamefont{M.~D.} \bibnamefont{Lukin}},
  \bibinfo{author}{\bibfnamefont{A.}~\bibnamefont{Yacoby}}, \bibnamefont{and}
  \bibinfo{author}{\bibfnamefont{R.}~\bibnamefont{Walsworth}},
  \textbf{\bibinfo{volume}{3}}, \bibinfo{pages}{858 EP }
  (\bibinfo{year}{2012}), \urlprefix\url{http://dx.doi.org/10.1038/ncomms1856}.

\bibitem[{\citenamefont{Hulet and Kleppner}(1983)}]{Hulet1983}
\bibinfo{author}{\bibfnamefont{R.~G.} \bibnamefont{Hulet}} \bibnamefont{and}
  \bibinfo{author}{\bibfnamefont{D.}~\bibnamefont{Kleppner}},
  \bibinfo{journal}{Phys. Rev. Lett.} \textbf{\bibinfo{volume}{51}},
  \bibinfo{pages}{1430} (\bibinfo{year}{1983}).

\bibitem[{\citenamefont{Signoles et~al.}(2017)\citenamefont{Signoles, Dietsche,
  Facon, Grosso, Haroche, Raimond, Brune, and Gleyzes}}]{Signoles2017}
\bibinfo{author}{\bibfnamefont{A.}~\bibnamefont{Signoles}},
  \bibinfo{author}{\bibfnamefont{E.~K.} \bibnamefont{Dietsche}},
  \bibinfo{author}{\bibfnamefont{A.}~\bibnamefont{Facon}},
  \bibinfo{author}{\bibfnamefont{D.}~\bibnamefont{Grosso}},
  \bibinfo{author}{\bibfnamefont{S.}~\bibnamefont{Haroche}},
  \bibinfo{author}{\bibfnamefont{J.~M.} \bibnamefont{Raimond}},
  \bibinfo{author}{\bibfnamefont{M.}~\bibnamefont{Brune}}, \bibnamefont{and}
  \bibinfo{author}{\bibfnamefont{S.}~\bibnamefont{Gleyzes}},
  \bibinfo{journal}{Phys. Rev. Lett.} \textbf{\bibinfo{volume}{118}},
  \bibinfo{pages}{253603} (\bibinfo{year}{2017}),
  \urlprefix\url{https://link.aps.org/doi/10.1103/PhysRevLett.118.253603}.

\bibitem[{\citenamefont{{See Supplemental Material at [URL will be inserted by
  publisher]}}(2018)}]{supplementary}
\bibinfo{author}{\bibnamefont{{See Supplemental Material at [URL will be
  inserted by publisher]}}} (\bibinfo{year}{2018}).

\bibitem[{\citenamefont{Nguyen et~al.}(2018)\citenamefont{Nguyen, Raimond,
  Sayrin, Corti\~nas, Cantat-Moltrecht, Assemat, Dotsenko, Gleyzes, Haroche,
  Roux et~al.}}]{ENS_CIRCSIM18}
\bibinfo{author}{\bibfnamefont{T.~L.} \bibnamefont{Nguyen}},
  \bibinfo{author}{\bibfnamefont{J.~M.} \bibnamefont{Raimond}},
  \bibinfo{author}{\bibfnamefont{C.}~\bibnamefont{Sayrin}},
  \bibinfo{author}{\bibfnamefont{R.}~\bibnamefont{Corti\~nas}},
  \bibinfo{author}{\bibfnamefont{T.}~\bibnamefont{Cantat-Moltrecht}},
  \bibinfo{author}{\bibfnamefont{F.}~\bibnamefont{Assemat}},
  \bibinfo{author}{\bibfnamefont{I.}~\bibnamefont{Dotsenko}},
  \bibinfo{author}{\bibfnamefont{S.}~\bibnamefont{Gleyzes}},
  \bibinfo{author}{\bibfnamefont{S.}~\bibnamefont{Haroche}},
  \bibinfo{author}{\bibfnamefont{G.}~\bibnamefont{Roux}}, \bibnamefont{et~al.},
  \bibinfo{journal}{Phys. Rev. X} \textbf{\bibinfo{volume}{8}},
  \bibinfo{pages}{011032} (\bibinfo{year}{2018}).

\bibitem[{\citenamefont{Brownnutt et~al.}(2015)\citenamefont{Brownnutt, Kumph,
  Rabl, and Blatt}}]{ION_BLATTRMP15}
\bibinfo{author}{\bibfnamefont{M.}~\bibnamefont{Brownnutt}},
  \bibinfo{author}{\bibfnamefont{M.}~\bibnamefont{Kumph}},
  \bibinfo{author}{\bibfnamefont{P.}~\bibnamefont{Rabl}}, \bibnamefont{and}
  \bibinfo{author}{\bibfnamefont{R.}~\bibnamefont{Blatt}},
  \bibinfo{journal}{Rev. Mod. Phys.} \textbf{\bibinfo{volume}{87}},
  \bibinfo{pages}{1419} (\bibinfo{year}{2015}).

\end{thebibliography}
\end{document}